\documentclass[superscriptaddress,twocolumn,showpacs,prl,floatfix]{revtex4}

\usepackage{color}
\usepackage{tabularx}
\usepackage{epsfig}
\usepackage{amsmath}

\usepackage{graphicx}

\begin{document}

\title{Critical exponents of the binomial Ising Spin Glass in
  dimension four; non-universality}

\author{P. H.~Lundow} 
\affiliation {Department of Mathematics and Mathematical Statistics,
  Ume{\aa} University, SE-901 87 Ume{\aa}, Sweden}

\author{I. A.~Campbell}
\affiliation{Laboratoire Charles Coulomb, Universit\'e Montpellier II,
  34095 Montpellier, France}

\begin{abstract}
Extensive simulations are made on the bimodal Ising Spin Glass (ISG) in dimension four.
The transition temperature is established using a combination of standard finite size
scaling and of thermodynamic derivative peak data.
Measurements in the thermodynamic limit regime are analysed so as to estimate
critical exponents and confluent correction terms. Comparisons with results on other
$4$d ISGs show that the  susceptibility
and correlation length critical exponents $\gamma$ and $\nu$ depend on the form of the
interaction distribution. From this observation it can be deduced that
critical exponents are not universal in ISGs.

\end{abstract}


\maketitle

\section{Introduction}

The universality of critical exponents is an important and remarkably
elegant property of standard second order transitions, which has been
explored in great detail through the Renormalization Group Theory
(RGT).  The universality hypothesis states that for all systems within
a universality class the critical exponents are rigorously identical
and do not depend on the microscopic parameters of the model. However,
universality is not strictly universal; there are known \lq\lq
eccentric\rq\rq models which are exceptions and violate the
universality rule in the sense that their critical exponents vary
continuously as functions of a control variable. The most famous
example is the eight vertex model solved exactly by Baxter
\cite{baxter:71}; there are other scattered cases, all in dimension
two as far as we are aware.

For Ising Spin Glasses (ISGs), the form of the interaction
distribution is a microscopic control parameter. It has been assumed
tacitly or explicitly that the members of the ISG family of
transitions obey standard universality rules, following the generally
accepted statement that \lq\lq Empirically, one finds that all systems
in nature belong to one of a comparatively small number of
universality classes\rq\rq \cite{stanley:99}.  However, we know of no
formal proof that universality must hold in ISGs; it was found thirty
years ago that the $\epsilon$-expansion for the critical exponents
\cite{gardner:84} in ISGs is not predictive since the first few orders
have a non-convergent behavior and higher orders are not known. This
can be taken as an indication that a fundamentally different
theoretical approach is required for spin glass transitions. Indeed
"Classical tools of RG analysis are not suitable for spin glasses"
\cite{parisi:01,castellana:11,angelini:13}.

ISG transition simulations are much more demanding numerically than
are those on, say, pure ferromagnet transitions with no interaction
disorder. The traditional approach in ISGs has been to study the
temperature and size dependence of observables in the near-transition
region and to estimate the critical temperature and exponents through
finite size scaling relations after taking means over large numbers of
samples. Finite size corrections to scaling should be allowed for
explicitly which can be delicate.  From numerical data, claims of
universality have been made repeatedly for ISGs
\cite{bhatt:88,katzgraber:06,hasenbusch:08,jorg:08} even though the
estimates of the critical exponents are very sensitive to the precise
value of the critical temperature and have varied over the years (see
Ref.~\cite{katzgraber:06} for a tabulation of historic estimates).

We have estimated the critical exponents of the bimodal ISG in
dimension $4$ using complementary strategies. First we use the
standard finite size crossing points of the Binder cumulant and other
phenomenological couplings to obtain estimates for the critical
temperature $\beta_c$ through finite size scaling \cite{notebeta}. We
also register the size dependence of the peaks of thermodynamic
derivatives which give independent estimates for $\beta_{c}$ and for
$\nu$ \cite{ferrenberg:91,weigel:09}.


We finally measure the temperature dependence of the thermodynamic
limit (ThL) ISG susceptibility $\chi(\beta,\infty)$ and second moment
correlation length $\xi(\beta,\infty)$ over a wide temperature
range. Using the scaling variable and scaling expressions appropriate
for ISGs as cited below \cite{daboul:04,campbell:06} together with the
optimal $\beta_{c}$ from the above measurements, we estimate the
critical exponents and the confluent corrections to scaling
\cite{wegner:72} from data taken over almost the entire paramagnetic
temperature range.

The numerical data on different ISGs in dimension $4$ show
conclusively that the critical exponents depend on the form of the
interaction distribution. It is relevant that it has been shown
experimentally that in Heisenberg spin glasses the critical exponents
depend on the strength of the Dzyaloshinski-Moriya interaction
\cite{campbell:10}.


\section{Ising Spin Glass simulations}

The Hamiltonian is as usual
\begin{equation}
  \mathcal{H}= - \sum_{ij}J_{ij}S_{i}S_{j}
  \label{ham}
\end{equation}
with the near neighbor symmetric distributions normalized to $\langle
J_{ij}^2\rangle=1$. The Ising spins live on simple hyper-cubic
lattices with periodic boundary conditions.  We have studied
bimodal ($\pm J$), Gaussian, and Laplacian distributions in $4$d.
Here we will discuss the bimodal ISG and will compare with published
measurements on two other $4$d ISGs \cite{jorg:08}.



The simulations were carried out using the exchange Monte-Carlo method for equilibration, on
$512$ individual samples at each size.  Data were registered after
equilibration for the energy $E(\beta,L)$, correlation length
$\xi(\beta,L)$, for the spin overlap moments $\langle |q|\rangle$,
$\langle q^2\rangle$, $\langle |q^3|\rangle$, $\langle q^4\rangle$,
and for the link overlap $q_{\ell}$ moments.  In addition the
correlations between the energy and certain observables $\langle
E\,U\rangle$ were also registered so that thermodynamic derivatives
could be evaluated using the relation $\partial U/\partial \beta =
\langle U\,E\rangle-\langle U \rangle\langle E\rangle$ where $E$ is
the energy \cite{ferrenberg:91}.  Bootstrap analyses of the errors in
the derivatives as well as in the observables themselves were carried
out.

For the present analysis we have observed the behavior of various
"phenomenological couplings", not only the familiar Binder cumulant
and correlation length ratio $\xi(\beta,L)/L$ but also other
observables showing critical behavior such as the kurtosis of the spin
overlap distribution, the kurtosis of the absolute spin overlap
distribution, and the variance and kurtosis of the link overlap
distribution.  Only part of these data are reported here.

Near criticality in a ferromagnet the heights of the peaks of the
thermodynamic derivative of many observables $\partial
U(\beta,L)/\partial \beta$ scale for large $L$ as
\cite{ferrenberg:91,weigel:09}
\[
\lbrack\partial U(\beta,L)/\partial \beta\rbrack_{\max} \propto L^{1/\nu}
\left(1+ b\, L^{-\omega/\nu}\right)
\]
and the temperature location of the derivative peak $\beta_{\max}(L)$
scales as $\beta_{c}-\beta_{\max}(L) \propto L^{-1/\nu}
\left(1+b'\,L^{-\omega/\nu}\right)$ The observables used for $U(\beta,L)$
\cite{ferrenberg:91} can be for instance the Binder cumulant
$g(\beta,L) =(3-\langle q^4\rangle/\langle q^2\rangle^2)/2$, the
logarithm of the finite size susceptibility $\ln(\chi(\beta,L))$, or
the logarithm of the absolute value of the spin overlap
$\ln(|q|(\beta,L))$.  Each of these data sets can give independent
estimates of $\nu$ and $\beta_c$ without any initial knowledge of
either parameter.

For the present analysis we note that both the minimum of the inverse
derivative $\lbrack\partial\beta/\partial U(\beta,L)\rbrack_{\min}$ and
the temperature location difference $\beta_{c}-\beta_{\min}(L)$ are
proportional to $L^{-1/\nu}$ to leading order. Hence
$\lbrack\partial\beta/\partial U(\beta,L)\rbrack_{\min}$ plotted
against $\beta_{\min}(L)$ with $L$ as an implicit variable must tend
linearly to an intercept $\lbrack\partial\beta/\partial
U(\beta,L)\rbrack_{\min}=0$ at $\beta_{\min} \equiv \beta_c$ for large
$L$.  All $\lbrack\partial\beta/\partial U(\beta,L)\rbrack_{\min}$
against $\beta_{\min}(L)$ plots should extrapolate consistently to the
true $\beta_c$.

Turning to spin glasses, for ISGs with symmetric interaction
distributions and a non-zero $\beta_c$ a general natural scaling
variable is $\tau = 1-(\beta/\beta_{c})^2$ ($w =
1-(\tanh(\beta)/\tanh(\beta_{c}))^2$ is also suitable for the bimodal
case) \cite{singh:86,daboul:04,campbell:06}.

In the ISG context $\beta^2$ replaces $\beta$ in the thermodynamic
derivative scaling rules but otherwise the same methodology can be
used as in the ferromagnet.  The thermodynamic derivative analysis,
which as far as we are aware has not been used previously in spin
glasses, provides reliable and precise estimates for $\beta_{c}^2$.
These $\beta_{c}^2$ estimates are consistent with those from the
traditional crossing point approach; which method has the least
sensitivity to finite size corrections depends on the individual
system.

The ThL SG susceptibility $\chi(\tau)$ including the leading
nonanalytic confluent correction term \cite{wegner:72} can be written
\begin{equation}
  \chi(\beta)= C_{\chi}\tau^{-\gamma}\left(1+a_{\chi}\tau^{\theta}+\cdots\right)
  \label{chiweg}
\end{equation}
where $\gamma$ is the critical exponent and $\theta$ the Wegner
non-analytic correction exponent, both of which are characteristic of
a universality class.  Following a protocol well-established in
ferromagnets \cite{kouvel:64,butera:02} one can define a temperature
dependent effective ThL exponent $\gamma(\beta)=
-\partial\ln\chi(\beta)/\partial\ln\tau$.  $\gamma(\beta)$ tends to
the critical $\gamma$ as $\beta^{2} \to \beta_{c}^2$ and to
$2d\beta_{c}^2$ as $\beta^{2} \to 0$ in simple [hyper]-cubic lattices.


As long as samples of finite size $L$ are in the ThL regime,
$\chi(\beta,L)$, $\xi(\beta.L)$ and other observables are independent
of $L$.  Working in the ThL has a number of advantages: the
temperatures studied are higher than the critical temperature so
equilibration is facilitated, the sample to sample variations are
automatically much weaker than at criticality, and there are no finite
size scaling corrections to take into account although the confluent
correction terms must be allowed for.  In ferromagnets the ThL
susceptibility and correlation length data can be fitted accurately
over the entire paramagnetic temperature range
\cite{campbell:08,campbell:11,lundow:11} by including just one further
effective correction term $k\tau^{\lambda}$ beyond the leading
non-analytic term, which bundles together all the higher order
correction terms.  We use this approximation also in the ISGs.  Hence
\begin{equation}
\gamma{\beta}= \gamma - \left(a_{\chi}\theta\tau^{\theta} +k\lambda\tau^{\lambda}\right)/
\left(1+a_{\chi}\tau^{\theta} +k\tau^{\lambda}\right)
\label{gamweg}
\end{equation}

A very effective method for analysing the ThL susceptibility data is
to plot $y = \partial\beta^2/\partial\ln\chi(\beta)$ against $x =
\beta^2$. With the two correction terms the expression used to fit the
ThL regime data is:
\begin{equation}
  \frac{\partial\beta^2}{\ln\chi(\beta)} = \frac{(\beta^2-\beta_c^2)(1+a_{\chi}\tau^{\theta}+k_{\chi}\tau^{\lambda_{\chi}})}
{\gamma + (\gamma-\theta) a_{\chi} \tau^{\theta}+(\gamma-\lambda_{\chi})k_{\chi}\tau^{\lambda_{\chi}}}
  \label{dbsqdlns}
\end{equation}
The critical intercept $y=0$ occurs when $x=\beta_{c}^2$, and the
initial slope starting at the intercept is $\partial y/\partial x
=-1/\gamma$.

The analogous natural scaling expression for the ISG second moment
correlation length $\xi(\beta)$ is \cite{campbell:06}
\begin{equation}
  \xi(\beta)/\beta = C_{\xi}\tau^{-\nu}\left(1+a_{\xi}\tau^{\theta}+k_{\xi}\tau^{\lambda}\right)
  \label{xiweg}
\end{equation}
with a temperature dependent effective exponent defined as $\nu(\beta)
= -\partial\ln(\xi(\beta)/\beta)/\partial\ln\tau$. The reason for the
factor $1/\beta$ arises from the generic form of the ISG $\xi(\beta)$
high temperature series \cite{campbell:06}. The $\beta=0$ limit in
ISGs in simple hyper-cubic lattices of dimension $d$ is $\nu(\beta=0)=
(d-K/3)\beta_c^2$ where $K$ is the kurtosis of the interaction
distribution.
The derivative corresponding to Eq.~\eqref{dbsqdlns} takes the form
\begin{equation}
  \frac{\partial\beta^2}{\ln(\xi(\beta)/\beta)} = \frac{\beta_c^2\tau(1+a_{\xi}\tau^{\theta})}{\nu + (\nu-\theta) a_{\xi} \tau^{\theta}}
  \label{dbsqdlnTxi}
\end{equation}
with the same $\beta_c^2$ and $\theta$ as for $\chi(\beta)$. The $y=0$
intercept is again $x=\beta_c^2$, with an initial slope at the
intercept equal to $\partial y/\partial x=-1/\nu$.

\section{ISG transitions in dimension $4$}

\begin{figure}
  \includegraphics[width=3.5in]{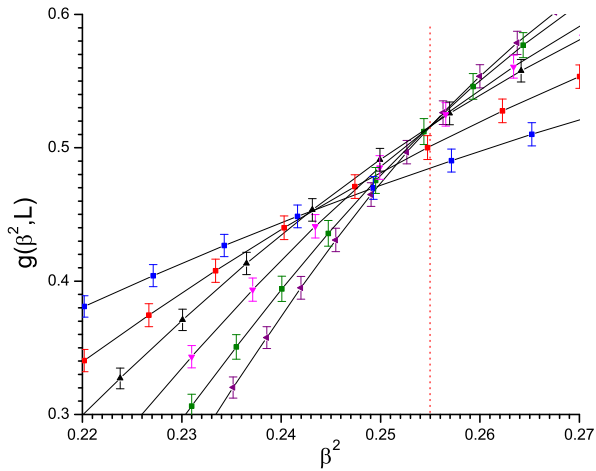}
  \caption{(Color online) The Binder cumulant $g(\beta,L)$ for even
    $L$ $4$d bimodal interaction samples. Symbol coding: blue squares
    $L=4$, red circles $L=6$, black triangles $L=8$, down triangles
    $L=10$, olive diamonds $L=12$, purple left triangles $L=14$.  The
    vertical red line corresponds to
    $\beta_{c}=0.505$.}\protect\label{fig:1}
\end{figure}

\begin{figure}
  \includegraphics[width=3.5in]{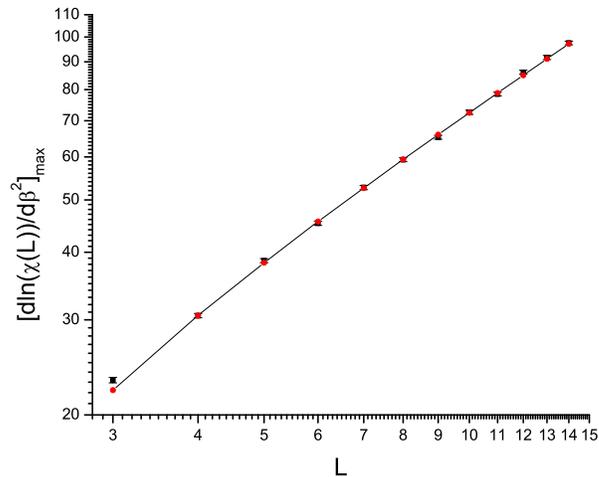}
  \caption{(Color online) The $4$d bimodal ISG thermodynamic
    derivative peak height $[\partial \ln(\chi(L)) \partial
      \beta^2]_{\max}$ as a function of size $L$. Black squares :
    measured, red circles : fit.}\protect\label{fig:2}
\end{figure}

\begin{figure}
  \includegraphics[width=3.5in]{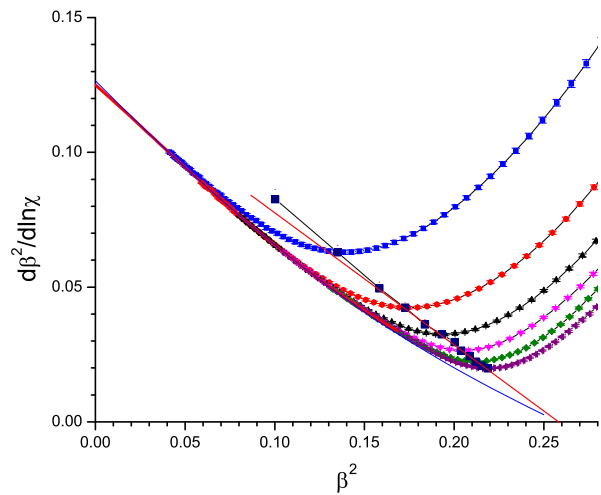}
  \caption{(Color online) $\partial\beta^2/\partial\ln\chi(\beta)$ for
    $4$d bimodal interaction samples. Even $L$ data only are shown to
    avoid clutter.  Symbol coding as in Fig.~\ref{fig:1}.  Large navy
    squares are the minima locations, odd and even $L$.  The full red
    curve is the the ThL calculated directly from HTSE
    \cite{daboul:04}.  Blue curve: fit
    Eq.~\eqref{dbsqdlns}.}\protect\label{fig:3}
\end{figure}

\begin{figure}
  \includegraphics[width=3.5in]{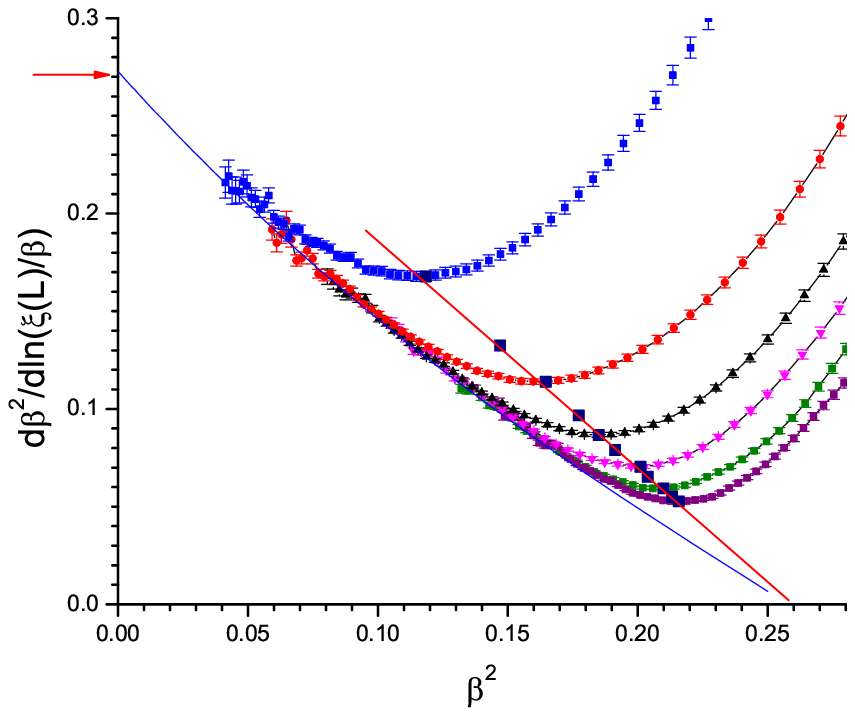}
  \caption{(Color online)
    $\partial\beta^2/\partial\ln(\xi(\beta)/\beta)$ for $4$d bimodal
    interaction samples. Even $L$ data only are shown to avoid
    clutter.  Symbol coding as in Fig.~\ref{fig:1}.  Large navy
    squares are the minima locations, odd and even $L$.  Blue curve:
    fit Eq.~\eqref{dbsqdlnTxi}.}\protect\label{fig:4}
\end{figure}

\begin{figure}
  \includegraphics[width=3.5in]{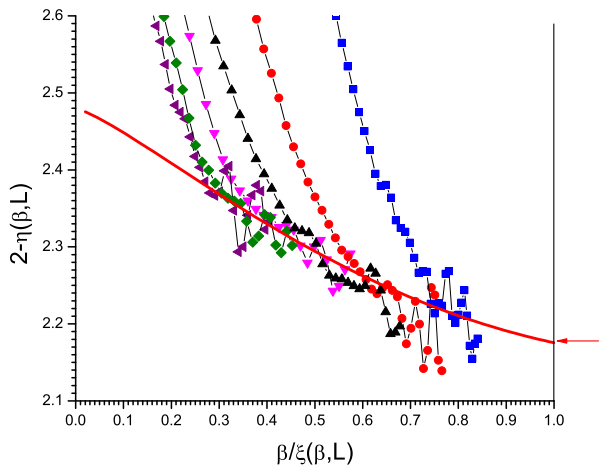}
  \caption{(Color online) The effective exponent $2-\eta(\beta,L)$
    Eq.~\eqref{meta} for $4$d bimodal interaction samples. Even $L$
    data only are shown to avoid clutter.  Symbol coding as in
    Fig.~\ref{fig:1}.  Red curve: fit through the ThL regime data.
    Red arrow : exact high temperature limit. The estimate of
    Ref.~\cite{banos:12} for the critical value $2-\eta(\beta_{c})$ is
    $2.320(13)$.}\protect\label{fig:5}
\end{figure}

High precision simulation measurements have been published on the $4$d
Gaussian ISG, and on a $4$d bimodal ISG with diluted interactions
($65\%$ of the interactions having $J=0$) \cite{jorg:08}.  The
critical temperature for the 4d Gaussian ISG was estimated from Binder
parameter and correlation length ratio measurements to be $\beta_{c}^2
= 0.307(3)$ in full agreement with earlier simulation estimates
$0.308(3)$ \cite{parisi:96,ney:98} and with the HTSE estimate
$\beta_{c}^2 = 0.314(4)$.  The simulations gave essentially identical
exponents for the two systems $\nu =1.02(2)$ $\eta = -0.275(25)$ so
indirectly $\gamma = 2.32(8)$.  Present data on the $4$d Gaussian (not
shown) using the thermal derivative analysis as above lead to a
$\beta_c^2$ in full agreement with that of Ref~\cite{jorg:08}, and
$\gamma= 2.36(3)$, with very weak corrections.
It seems very reasonable to assume that if the present procedure were
applied to the diluted bimodal system, it would confirm the
conclusions of Ref.~\cite{jorg:08} for that system also. It can be
noted that for this system the finite size correction to scaling in
the Binder cumulant is so small as to be unobservable.

For the $4$d bimodal ISG the HTSE critical temperature and exponent
estimates are \cite{daboul:04} $\beta_{c}^2 = 0.26(2)$, $\gamma =
2.5(3)$, and $\theta \sim 1.5$.  From extensive domain wall free
energy measurements to $L = 10$ Hukushima gave an estimate
$\beta_{c}^2 = 0.25(1)$ \cite{hukushima:99}.  Inspection of the raw
data show strong finite size corrections; extrapolation to larger $L$
leads to an infinite size limit definitely greater than $0.25$.

From early simulation measurements up to $L = 10$ a critical
temperature $\beta_{c}^2 = 0.243(7)$ was estimated \cite{marinari:99}
using the Binder parameter crossing point criterion. However, finite
size corrections to scaling were not allowed for. Recent simulations
up to $L=16$ \cite{banos:12} show large $L$ Binder cumulant crossings
up to $\beta^2 =0.252$. Our present Binder cumulant data up to $L=14$
show very similar results, Figure 1.  In this figure crossing points
for the largest $L$ can be seen to cluster around $\beta^2 =0.255$,
which provides a lower limit on $\beta_{c}^2$.

In addition, it can be seen that the value of the cumulant at the
present high $L$ crossing points is $g_{cross}(\beta^2) = 0.520(3)$,
which is a lower limit on the infinite size critical
$g_{c}(\beta_{c}^2)$.  In comparison $g_{c}(\beta_{c}^2)$ has been
estimated at $0.470(5)$ and $0.472(2)$ for the $4$d Gaussian and
diluted bimodal ISG respectively \cite{jorg:08}. The bimodal critical
crossing point is thus 25 standard deviations above the diluted
bimodal critical value. As the critical $g_c(\beta_{c}^2)$ is, for a
given geometry, a parameter characteristic of a universality class,
the bimodal and diluted bimodal systems are not in the same
universality class; neither are the bimodal and the Gaussian ISGs.

We will now turn to the inverse susceptibility derivative data,
Figures 2, 3 and 4. In Figure 2 the thermodynamic derivative peak
height $y(L) = [\partial \ln(\chi(\beta^2,L))/ \partial
  \beta^2]_{\max}$ is fitted by $y(L) =
11.0L^{1/\nu}(1-0.95L^{\theta/\nu})$ with $\nu = 1.20(2)$ and $\theta$
fixed at $1.75$, from ThL data discussed below. The value of the
estimate for $\nu$ requires no information on the critical temperature
$\beta_{c}$. It is significantly higher than the estimate $\nu =
1.068(7)$ given by Ref.~\cite{banos:12}.

In Figure 3 the $[\partial \beta^2/\partial\ln\chi(\beta^2,L)]_{\min}$
points for different $L$ have a straight line limit at large $L$ which
tends to an intercept at $x = 0.258(1)$.  Derivative minima plots of
the same type for other observables (not shown) confirm this value for
$\beta_{c}^2$.  These estimates are very reliable as they come from
parameter free straight line limit fits. The estimate of
Ref.~\cite{banos:12} is $\beta_{c}^2 = 0.2523(6)$; this value is
sensitive to the estimate for the correction to scaling exponent.

The curves for individual $L$ and the HTSE curve all lie on a size
independent ThL envelope curve to the left of the figure.  A
satisfactory fit passing through all the ThL data and the critical
point can be made with a single correction term only.  The optimal fit
parameters are $\gamma = 3.05(5)$, $\theta = 1.75(5)$, $a_{\chi} =
1.40(3)$.  A similar analysis made on the correlation length data,
Figure 4, provides a completely consistent estimate for $\beta_{c}^2$
from the linear variation of the minima points. The ThL data are
fitted with the parameters $\nu =1.20(3)$ and $a_{\xi}=0.17(2)$
together with the same $\theta = 1.75$ as for the susceptibility fit.

An estimate of the exponent $\eta$ can be made from a plot of
\begin{equation}
2 -\eta(\beta,L) = \partial \ln(\chi(\beta,L)) / \partial
\ln(\xi(\beta,L)/\beta)
\label{meta}
\end{equation} 
against $\beta/\xi(\beta,L)$, Figure 5. Extrapolating the ThL regime
data to criticality at $\beta/\xi(\beta,L)=0$ leads to a direct
estimate $\eta = -0.48(3)$ without needing any assumption concerning
$\beta_{c}^2$ or finite size corrections. The data clearly show that
the estimate $\eta = -0.320(13)$ of Ref.~\cite{banos:12} is low.

The present exponents are reliable and much more accurate than the
HTSE estimates principally because the uncertainty in $\beta_{c}^2$ is
reduced by a factor of more than $10$ thanks principally to the
thermal derivative peak simulation data.  The exponents can be
compared to the values found \cite{jorg:08} for the $4$d Gaussian and
diluted bimodal systems which were almost identical to each other :
$\gamma=2.32(8)$ and $\nu= 1.02(2)$ for the Gaussian and $\gamma
=2.33(6)$, $\nu = 1.025(15)$ for the diluted bimodal.  The critical
exponents of the $4$d bimodal ISG are quite different from those of
the $4$d Gaussian and diluted bimodal ISGs.

\section{Conclusions}

Simulations on the $4$d bimodal ISG up to size $L=14$ provide
numerical data on finite size scaling observables, on the ISG
susceptibility and on the correlation length.  The critical
temperature $\beta_{c}$ derived from the simulation data using a
thermodynamic derivative technique \cite{ferrenberg:91} is in full
agreement with, but is considerably more precise than, the estimate
from HTSE alone \cite{daboul:04}. Because of the analysis techniques
used it is also more reliable than previous numerical estimates. Data
in the thermodynamic limit regime were analysed to obtain considerably
improved critical exponent $\gamma$, $\nu$ and $\theta$ estimates
together with the strengths of leading confluent correction terms.

The accurate estimates of $\gamma$ and $\nu$ and the critical value of
the Binder cumulant show that the $4$d bimodal ISG is in a different
universality class from the $4$d Gaussian or diluted bimodal ISGs
\cite{jorg:08}. Other results on ISGs in dimension $4$ \cite{lundow}
and in dimension $5$ \cite{lundow:13a} confirm that spin glasses with
different interaction distributions have different critical
exponents. These results clearly demonstrate that the standard RGT
universality rules do not apply in ISGs.

\section{Acknowledgements}
We are very grateful to Koji Hukushima for comments and communication
of unpublished data. We thank Amnon Aharony for constructive
criticism. The computations were performed on resources provided by
the Swedish National Infrastructure for Computing (SNIC) at the High
Performance Computing Center North (HPC2N).


\begin{thebibliography}{20}
\bibitem{notebeta} We will use inverse temperatures $\beta = 1/T$ throughout.
\bibitem{gauss13} The Gaussian series was calculated to $13$ terms but explicit evaluation shows that
there is a numerical error in the final term.
\end{thebibliography}
\end{document}